\documentclass[preprint,12pt]{elsarticle}

\usepackage{xurl}

\usepackage{amssymb}
\usepackage{multirow}
\usepackage{arydshln}
\usepackage[table]{xcolor}
\definecolor{light-gray}{gray}{0.9}

\usepackage{color, colortbl}
\usepackage{amsfonts} 

\usepackage{amssymb}

\RequirePackage{amsthm,amsmath}

\usepackage{breqn}

\usepackage{algorithmic}

\usepackage{amsmath}
\usepackage{array}
\usepackage{makecell}
\usepackage{mathtools}

\makeatletter
\newcommand{\thickhline}{%
    \noalign {\ifnum 0=`}\fi \hrule height 1pt
    \futurelet \reserved@a \@xhline}

\usepackage{algorithm}
\usepackage{algorithmic}

\makeatletter
\newcommand\fs@norules{\def\@fs@cfont{\bfseries}\let\@fs@capt\floatc@ruled
  \def\@fs@pre{}%
  \def\@fs@post{}%
  \def\@fs@mid{\kern3pt}%
  \let\@fs@iftopcapt\iftrue}
\makeatother
\floatstyle{norules}
\restylefloat{algorithm}

\journal{xxx}

\begin{document}

\begin{frontmatter}



\title{AIRU-WRF: A Physics-Guided Spatio-Temporal Wind Forecasting Model and its Application to the U.S. {Mid} Atlantic Offshore Wind Energy Areas}


\author[1]{Feng Ye}
\author[2]{Joseph Brodie}
\author[3]{Travis Miles}
\author[1]{Ahmed Aziz Ezzat}

\affiliation[1]{organization={Industrial \& Systems Engineering, Rutgers University},
            city={Piscataway},
            postcode={08854},
            state={NJ},
            country={USA}}

\affiliation[2]{organization={AKRF Inc.},
            city={New York},
            postcode={10016},
            state={NY},
            country={USA}} 
            
\affiliation[3]{organization={Marine \& Coastal Sciences, Rutgers University},
            city={Piscataway},
            postcode={08901},
            state={NJ},
            country={USA}}

\begin{abstract}
The reliable integration of wind energy into modern-day electricity systems heavily relies on accurate short-term wind forecasts. We propose a spatio-temporal model called AIRU-WRF (short for the \underline{AI}-powered \underline{R}utgers \underline{U}niversity \underline{W}eather \underline{R}esearch \& \underline{F}orecasting), which {combines} numerical weather predictions (NWPs) with local observations in order to make wind speed forecasts that are short-term (minutes to hours ahead), and of high resolution, both spatially (site-specific) and temporally (minute-level). In contrast to purely data-driven methods, 
we undertake a ``physics-guided'' machine learning approach which captures salient physical features of the local wind field \textit{without} the need to explicitly solve for those physics, including: (\textit{i}) modeling wind field advection and diffusion via physically meaningful kernel functions, (\textit{ii}) integrating exogenous predictors that are both meteorologically relevant and statistically significant; and (\textit{iii}) linking the multi-type NWP biases to their driving {mesoscale} weather conditions. 
Tested on real-world data from the U.S. {Mid} Atlantic where several offshore wind projects are in-development, AIRU-WRF achieves notable improvements, in terms of both wind speed and power, relative to various forecasting benchmarks including physics-based, hybrid, statistical, and deep learning methods. 

\end{abstract}




\end{frontmatter}


\section{Introduction}\label{intro}
Offshore wind (OSW) is one of the fastest growing sources of renewable energy worldwide. 
Several countries have set ambitious targets to {increase} the penetration of OSW into their electricity systems. For instance, the United States ({U.S.}) plans to install $30$ Gigawatts (GW) of OSW capacity by $2030$, distributed across five major geographical regions. Among those, the U.S. {Mid} Atlantic {and Northeastern U.S.} (the focus of this work) is set to contribute more than a third of the total planned capacity. To meet this target, several GW-scale OSW projects are currently under development in this region 
\cite{nyserdaroadmap2019,lease2021}. 

The reliable integration of those soon-to-be-operational OSW farms into the power grid will be contingent on accurate, high-resolution, short-term wind forecasts. To that end, we propose the \underline{AI}-powered \underline{R}utgers \underline{U}niversity \underline{W}eather \underline{R}esearch \& \underline{F}orecasting (AIRU-WRF) model, in order to make accurate OSW forecasts that are short-term (minutes to hours ahead), and of high resolution, both spatially (site-specific) and temporally (minute-level). Such hyper-detailed forecasts\textemdash practically unattainable by stand-alone physics-based models\textemdash are pivotal to several OSW operations, 
including power estimation \cite{lee2015power, nasery2023yaw}, 
economic dispatch and reserve planning \cite{jiang2021propagation, barry2022risk}, operations and maintenance \cite{petros2021, petros2022, papadopoulos2023joint}, among others. 

Obtaining high-resolution, short-term wind forecasts, however, is challenging, mainly due to the limitations of {mesoscale} numerical weather predictions (NWPs) at finer spatial and temporal scales. The inaccurate parameterization of sub-grid physical processes, spatial averaging assumptions, and local effects largely compromise the value of NWPs for (near) real-time operations. Figure \ref{figbias} shows two days of hourly wind speed forecasts from a state-of-the-art {mesoscale} NWP model, along with $10$-min co-located wind speed observations in the U.S. {Mid} Atlantic. When 
downscaled to the $10$-min, site-specific resolution at which the forecasts are {needed}, NWPs suffer from notable errors and biases, of multiple types, including temporal biases (early/late predictions) and shift biases (under- or over-prediction) \cite{sweeney2020future}.

\begin{figure}
\includegraphics[trim = .75cm .4cm 0 0,width=\textwidth]{Bias2.jpg}  
\centering  
\caption{Two days of meso-scale NWPs from the RU-WRF meso-scale model \cite{optis2020validation}, along with co-located wind speed observations. Both data and forecasts are obtained in proximity to the planned OSW energy areas in the NY/NJ Bight. Despite their value, NWPs often exhibit notable forecast biases, of multiple types and magnitudes, when directly downscaled to higher spatio-temporal resolutions.} 
\label{figbias} 
\end{figure}  

To remedy those limitations, 
statistical and machine learning (ML) methods have emerged as a powerful approach for high-resolution, short-term wind forecasting \cite{sweeney2020future, AzizRS}. One criticism often pointed at statistical and ML methods, however, is that they are, by and large, \textit{physics-agnostic}, i.e., they are formulated with little consideration of the physics of wind field formation and propagation, 
and hence, may be susceptible to model specifications that violate those first principles. This has driven an active area of research in ML referred to as \textit{physics-informed}, or \textit{physics-guided} learning. Physics-guided learning is broadly defined as the integration of physical principles, laws, or physics-based information within ML models in order to guide them to adhere to certain physical aspects underpinning the driving physical process. 

One direct approach to loosely inject physics within ML-based wind forecasting is by using NWP outputs as regressors to a statistical- or ML-based formulation. This approach is often referred to as a ``hybrid'' forecasting model as it attempts to calibrate the physics-based NWPs at a set of target locations and time resolutions by learning a functional mapping that adjusts future NWPs closer to incoming observations \cite{chen2013wind,dong2016wind,hu2021hybrid}. Our work broadly belongs to this family of hybrid approaches, but departs from the vast majority of the methods therein by bearing on physical knowledge to guide the selection of certain parameters, features, and statistical constructs in the ML-based forecasting model, rather than relying on a purely data-driven correction of NWP outputs. 
As such, our method does not seek to ``replace'' NWP models, 
but rather borrow strength across both the physical and statistical learning paradigms. 
In light of that, the unique aspects of this work are summarized as follows: 

($\mathcal{A}_1$) The vast majority of wind forecasting models either overlook the spatio-temporal dependence in wind fields, or at best, model it using physically irrelevant kernel functions which unrestrictedly learn the spatial and temporal correlations in the historical data. 
In contrast, AIRU-WRF adopts a special 
class of physically meaningful kernels, collectively dubbed as the ``Lagrangian reference framework,'' for which we select the parameters in part using NWPs. We show how this physically aligned covariance modeling approach encodes the first principles of wind field formation and propagation, and how it translates into significant forecast accuracy gains. 

(\emph{A2}) We connect the multi-type NWP biases (such as those shown in Figure \ref{figbias}) with their driving {mesoscale} weather conditions via a physically justifiable calibration model, wherein we construct exogenous predictors that are both physically meaningful (in terms of meteorological relevance) and statistically significant (in terms of explanatory power). 
We dynamically update this feature input set, yielding a parsimonious statistical representation that is shown to enhance the predictive quality of the final forecasts. 

(\emph{A3}) Unlike several methods that solely focus on single-valued (or point) forecasting, AIRU-WRF makes probabilistic predictions, even at locations where no or sparse data is available. 
We leverage this capability to generate wind field forecast ``maps'' (in the form of evolving two-dimensional images) for effective visualization and communication of the forecast outputs to OSW energy stakeholders and end-users. 

(\emph{A4}) In terms of practical relevance, we focus on the OSW regions in the U.S. {Mid} Atlantic where several GW-scale projects are under development. 
Our models, results, and analyses can potentially provide timely insights to the developers and operators of those soon-to-be-operational wind farms.

The remainder of this paper is organized as follows. Section \ref{data} describes the data used in this study and its relevance to the U.S. OSW wind industry. 
 Section \ref{method} introduces the building blocks of AIRU-WRF, followed by Section \ref{results} where forecast evaluations, results, and discussions are presented. Section \ref{conclusion} concludes the paper and highlights future research directions.

\section{Data Description} 
\label{data}
\begin{figure}  
\centering  
\includegraphics[trim = 0cm .75cm 0cm 0cm, width = \linewidth]{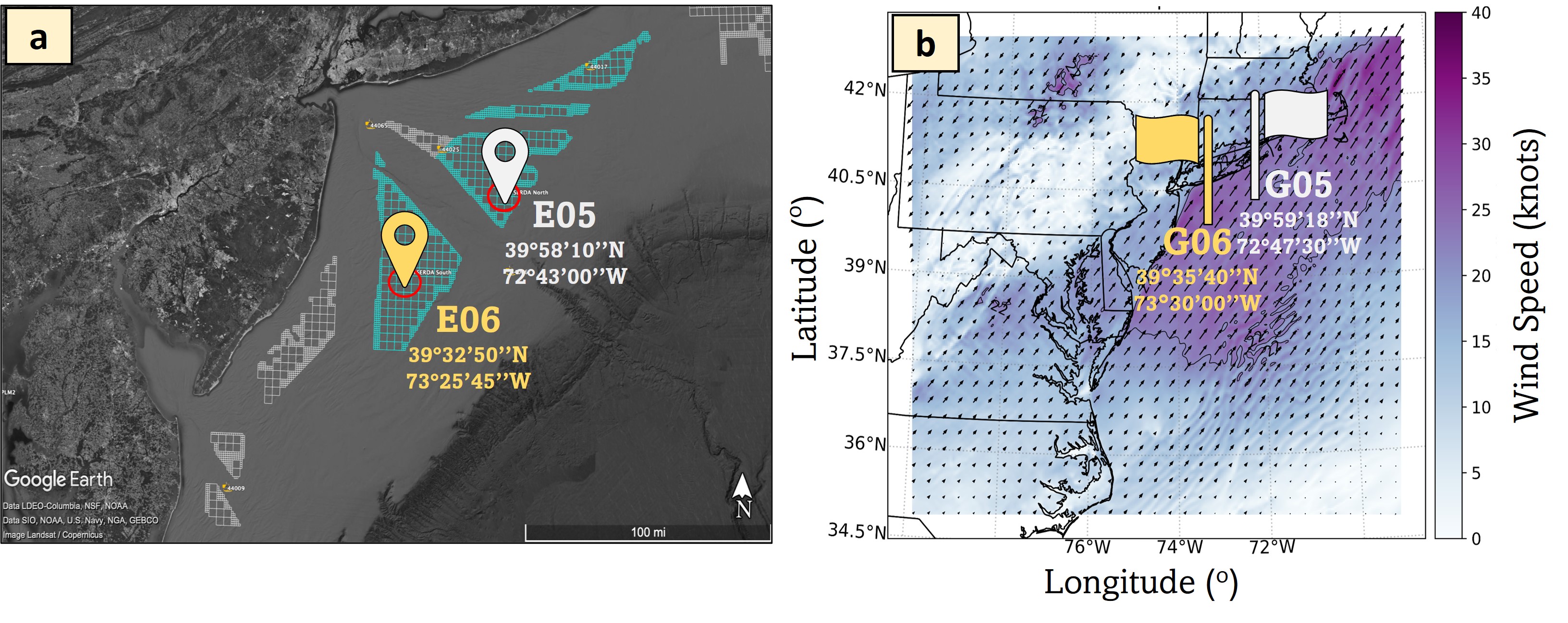}
\caption{(a) Locations of the two buoys, E05 and E06, in the NY/NJ Bight. Green and grey regions represent OSW energy planning and lease areas, respectively 
\cite{lease2021}; (b) Locations of the two closest RU-WRF grid points, G05 and G06, on top of the NWP forecasts on November 7th, 2019 at 18:00 GMT.}
\label{figlocation}  
\end{figure}

This work has been motivated by the ongoing large-scale OSW developments in the U.S. {Mid} Atlantic, and in particular, the {New York (NY)/New Jersey (NJ)} Bight (shown in Figure \ref{figlocation}) \cite{lease2021}. 
We make use of two sources of data from this geographical region, with varying spatial and temporal resolutions: (i) A set of hub-height wind speed observations collected by two floating Lidar buoys (E05 and E06); and (ii) a set of co-located NWP outputs, obtained via a state-of-the-art {mesoscale} meteorological model (RU-WRF). Details of both sets of data are described next. 

\subsection{Local observations from the NY/NJ Bight}\label{truedata}
A set of wind speed observations is obtained from two floating Lidar buoys: E05 Hudson North (E05) and E06 Hudson South (E06) 
\cite{measurement}. 
The two buoys are $\sim$\hspace{-0.05mm}77 {km} apart. 
The observations are recorded in $10$-min intervals. 
We focus on the $100$-m altitude, which is a common hub-height of typical wind turbines. The available data spans a total duration of 6 months, distributed into two distinct periods, representing winter and summer intervals, respectively. Specifically, the winter period spans from November 1st, 2019 to February 29th, 2020, while the summer period spans from May 5th to June 30th, 2020. 
Throughout both those periods, the prevailing wind direction was mostly westerly, with average wind speeds of $10.42$ m/s and $8.64$ m/s for the winter and summer periods, respectively, and average wind directions 
of $304.44^\circ$ 
and $234.98^\circ$, for the winter and summer periods, respectively. 
{Our choice of the study periods is attributed to the quality and completeness of the hub-height observational data during those times of the year, relative to other time periods where long streaks of missing observations prevented us from conducting extensive and reliable forecasting experiments. Our analysis of the correspondent year-round NWP data suggests that the wind conditions observed during the study period cover a wide spectrum of the meteorological regimes that typically manifest in this region.}

\subsection{Numerical Weather Predictions from RU-WRF}\label{ru-wrf}
The Rutgers University Center for Ocean Observing Leadership (RUCOOL) at Rutgers University runs a daily real-time version of WRF, called RU-WRF, which is tailored to the U.S. Mid Atlantic OSW energy areas \cite{dicopoulos2021weather}. {
RU-WRF runs a parent nest at $9$ km resolution out to $120$ hours and a child nest centered on the NJ shelf at $3$ km resolution out to $48$ hours, generating hourly forecasts of multiple meteorological variables, which are listed in Table \ref{tableNWP}. The model initial and boundary conditions are obtained from the National Weather Service (NWS) Global Forecasting System (GFS), with the model re-initialized from GFS daily at 00Z. The continuous model data archive from December 2019 to March 2023 \cite{RUCOOL} includes hourly output compiled from the 3-km domain model.} 

{RU-WRF has been recently validated by the National Renewable Energy Laboratory (NREL) \cite{optis2020validation}. The physics parameterizations and model setup in the version of RU-WRF used here are consistent with \cite{Mike2021}, however, the vertical levels have been increased from $40$ to $48$ to enhance resolution in the boundary layer near the sea surface, the surface layer scheme is now Mellor–Yamada–Nakanishi–Niino (MYNN) \cite{olson2021description}. The model is a non-data assimilative system, however, it does utilize a novel ocean sea surface temperature satellite product designed to capture the nearshore coastal upwelling \cite{MURPHY2021} at each daily initialization to provide an accurate bottom boundary condition within the Mid Atlantic Bight (MAB). Generally, the modeling philosophy is to leverage the highly accurate initial conditions of the NWS GFS model, but enhance local simulations through the use of higher horizontal and vertical resolutions and accurately represent the essential ocean feature of the MAB, the coastal upwelling.} 
The nearest NWP grid points to E05 and E06 are denoted as G05 and G06, respectively, and are shown in Figure \ref{figlocation}(b), on top of the NWPs for a select day and time in November, 2019. 


\begin{table}
\centering
\small
\caption{NWP variables, their descriptions, and units, as extracted from RU-WRF.}
\resizebox{\columnwidth}{!}{%
\begin{tabular}{l|c|c}
\hline
 \textbf{NWP variable}  & \textbf{Description} & \textbf{Unit} \\
\hline
WIND SPEED &       Wind speed forecast at $100$-m altitude                              &    m/s  \\	
SWDOWN     & 	    Surface downwelling shortwave Flux      &    W/m\textsuperscript{2}  \\
LWUPB      &    	Surface upwelling longwave flux         &    W/m\textsuperscript{2} \\  
GLW        &   	    Surface downwelling longwave flux       &    W/m\textsuperscript{2} \\
SNOWNC     &    	Accumulated total grid scale snow and ice  & mm \\
TEMP    &   
Sea surface temperature                 & K \\
DIF\_FRAC &
Diffuse fraction of surface shortwave irradiance   & - \\
LANDMASK    & 	    Land mask (1 For Land, 0 For Water)  & - \\
LAKEMASK   &   	    Lake mask (1 For Lake, 0 For Non-Lake)  & - \\
PBLH       &    	Height of the top of the planetary boundary layer (PBL)  & m \\
HUMIDITY   &  	    Relative humidity (surface level)   & $\%$  \\
PRESSURE   &     	Sea level pressure   & hPa\\
MDBZ      &    	    Maximum radar reflectivity   & dBZ\\
U         &    	    Eastward wind component at 100-m altitude                                 &    m/s \\
V         &     	Northward wind component at 100-m altitude                                   &    m/s\\
WINDGUST   &    	\makecell[c]{Wind gust, computed by mixing down momentum\\from the level at the top of the planetary boundary layer}                               &    m/s\\
\hline
\end{tabular}
}
\label{tableNWP}
\end{table}

\section{Methodology}\label{method}
Let $Y(\mathbf{s}, t)$ be the random variable denoting the hub-height, spatio-temporal wind speed, where $\mathbf{s} \in \mathbb{R}^2$ is a pair of spatial coordinates, and $t \in \mathbb{Z}^+$ denotes time such that $\mathbb{Z}^+$ is the set of non-negative integers. Similarly, let $\mathbf{X}(\mathbf{s}, t) = [X_1(\mathbf{s}, t), ..., X_p(\mathbf{s},t)]^T$ be a set of 
$p$ meteorological variables, other than wind speed, for which hourly NWP forecasts are available. Examples of such variables are the NWP outputs that are listed in Table \ref{tableNWP}. 

At our disposal are two sets of data with distinct spatial and temporal resolutions:  (\textit{i}) a set of hub-height, spatio-temporal wind speed observations denoted by 
$\mathbf{y}(\mathbf{s}, t) =[y(\mathbf{s}_1, t_1), ..., y(\mathbf{s}_n, t_c)]^T$, 
where $n$ is the number of measurement locations and $t_c$ denotes the current time, and (\textit{ii}) a set of correspondent NWP forecasts for $Y(\mathbf{s}, t)$, denoted by 
$\hat{\mathbf{y}}(\mathbf{s}, t) = [\hat{{y}}({{\mathbf{s}}}_{1},t_1), ..., \hat{{y}}({{\mathbf{s}}}_n, t_k)]^T$, 
as well as for $\mathbf{X}(\mathbf{s}, t)$, which are denoted by 
$\hat{\mathbf{x}}(\mathbf{s}, t)$. 
Note that (\textit{i}) $t_k > t_c$ because NWP forecasts are available for both the historical observations (up to $t_c$), as well as for future forecast horizons (beyond $t_c$); and (\textit{ii}) $\mathbf{y}$ and $\hat{\mathbf{y}}$ may not have the same temporal resolution (In our context, $10$-min for the actual data $\mathbf{y}$, while NWPs, $\hat{\mathbf{y}}$, are of hourly resolution). Hereinafter, notation for \textit{variables} will be in uppercase, while that for \textit{data} will be in lowercase. 

The overarching formulation of AIRU-WRF is shown in (\ref{eq-ourmodel}), where $\mu(\mathbf{s}, t)$ and $\eta(\mathbf{s}, t)$ are two independent spatio-temporal functions, serving two distinct purposes, while $\epsilon(\mathbf{s}, t)$ is the Gaussian white noise process. In specific, $\mu(\mathbf{s}, t)$ is intended to capture large-scale, low-frequency variations in the wind field that manifest themselves over relatively coarse time scales and spatial resolutions. We refer to this as the ``{mesoscale}.'' In contrast, $\eta(\mathbf{s}, t)$ characterizes the higher-resolution, site-specific variations that $\mu(\mathbf{s}, t)$ typically fails to capture. We refer to this as the ``local'' or ``sub-{mesoscale}'' variation. 
\begin{equation}\label{eq-ourmodel}
Y(\mathbf{s},t)= 
\hspace{-0.6cm} \underbrace{
\mu(\mathbf{s},t|\mathbf{y}, \hat{\mathbf{y}}, \hat{\mathbf{x}})
}_{\parbox{11em}{\scriptsize \centering
{Mesoscale} variation
%
}} \hspace{-0.75cm}
+ \hspace{-.9cm} 
\underbrace{
\eta(\mathbf{s},t|\mathbf{y}, \hat{\mathbf{x}})
}_{\parbox{11em}{\scriptsize \centering
Local variation
}}
\hspace{-0.9cm}
+ \hspace{-0.1cm} 
\underbrace{
\epsilon(\mathbf{s},t)
}_{\text{White noise}}.
\end{equation}
Next, we discuss the role and formulation of $\mu(\mathbf{s}, t)$ and $\eta(\mathbf{s}, t)$ in Sections \ref{sec:mu} and \ref{sec:eta}, respectively. 

\subsection{Physics-guided modeling of $\mu(\mathbf{s}, t)$}\label{sec:mu}
The role of $\mu(\mathbf{s},t)$ is to characterize the larger-scale variation in the wind field, which are mostly driven by physical phenomena that manifest themselves over relatively longer time scales (hours to days) and coarser spatial resolutions ({mesoscale}). Examples of those large-scale variations include trends, diurnal and semi-diurnal cycles, regime alternations, etc.  
Our assumption herein is that NWPs can play a key role in capturing such larger-scale fluctuations by virtue of their embedded physics. 
As such, we can think of $\mu(\mathbf{s}, t)$ as an NWP calibration, 
except that it is a physically motivated one. 

The first step of this calibration is to interpolate the hourly NWP variables $\hat{{Y}}(\mathbf{s}, t)$ and $\hat{\mathbf{X}}(\mathbf{s}, t)$ 
into the $10$-min resolution at which the forecasts are to be made, yielding the interpolated NWP variables $\tilde{{{Y}}}(\mathbf{s}, t)$ and $\tilde{{\mathbf{X}}}(\mathbf{s}, t)$, respectively. 
For interpolation of both sets, we use cubic splines. 
Let us then denote by $\tilde{\mathbf{G}}(\mathbf{s}, t) = [\tilde{G}_1(\mathbf{s}, t), ..., \tilde{G}_m(\mathbf{s},t)]^T$ 
the set of $m$ spatio-temporal explanatory variables, which are to be included as regressors in modeling 
$\mu(\mathbf{s}, t)$. For example, $\tilde{\mathbf{G}}(\mathbf{s}, t)$ may include the ``most informative'' subset of $\tilde{\mathbf{X}}(\mathbf{s}, t)$, in addition to other exogenous variables that possess some degree of explanatory power in calibrating NWPs. 
In light of that, 
we propose the following form for $\mu(\mathbf{s}, t)$:
\begin{equation}\label{mu}
\mu(\mathbf{s}, t)=
\underbrace{\mathbf{a}^{T}\tilde{\mathbf{{Y}}}^{\ell}(\mathbf{s},t)
+\mathbf{b}^{T}\tilde{\mathbf{{G}}}(\mathbf{s},t)}_{\text{additive bias correction}}+
\underbrace{
\mathbf{c}^{T}\tilde{\mathbf{G}}(\mathbf{s},t)
\tilde{{Y}}(\mathbf{s},t),}_{\text{multiplicative bias correction}}\ 
\end{equation}
where $\mathbf{a}$, $\mathbf{b}$ and $\mathbf{c}$ are sets of unknown parameters (to be estimated), and $\tilde{\mathbf{{Y}}}^{\ell}(\mathbf{s},t) = [\tilde{Y}(\mathbf{s}, t), ..., \tilde{Y}(\mathbf{s}, t - \ell)]^T$ denotes the 
set of (interpolated) lagged NWP forecasts of wind speed, up to lag $\ell$. The motivation behind (\ref{mu}) 
is to correct the multi-type biases of NWPs by linking them to their driving {mesoscale} weather conditions encoded in $\tilde{\mathbf{G}}(\mathbf{s}, t)$ and $\tilde{{\mathbf{Y}}}^{\ell}(\mathbf{s},t)$. 

The formulation in (\ref{mu}) targets two distinct types of NWP biases: Additive bias refers to the systematic NWP inaccuracies that can be corrected by scaling. Examples are shift biases (over- or under-prediction), temporal biases (early/late prediction), and spatial biases (location-dependent errors)\textemdash Recall Figure \ref{figbias}. 
The inclusion of lagged values in $\tilde{\mathbf{{Y}}}^{\ell}(\mathbf{s},t)$ is motivated by their potential role in especially correcting the temporal biases. 
The multiplicative bias term, on the other hand, addresses nonlinear biases that cannot be simply adjusted by scaling, such as regime-specific NWP errors 
(e.g., errors that are more pronounced at higher or lower winds).  

The next key question is to identify the variables constituting the elements of the set $\tilde{\mathbf{G}}(\mathbf{s}, t)$. Our approach to construct $\tilde{\mathbf{G}}(\mathbf{s}, t)$ is to include variables which are both physically meaningful (in terms of meteorological relevance) \textit{and} statistically significant (in terms of explanatory power), thus forming the basis of our physics-guided calibration. As a starter, $\tilde{\mathbf{G}}(\mathbf{s}, t)$ will include a subset of the NWP variables in $\tilde{\mathbf{X}}(\mathbf{s}, t)$ which are known to possess a meteorological association with wind field physics. Out of the NWP outputs of Table \ref{tableNWP}, {prior physical} knowledge suggests the potential inclusion of six features: air pressure, surface temperature, wind gust, relative humidity, eastward and northward wind components, since those variables are physically known to contribute, either directly or indirectly, to local wind field formation and propagation. From a purely statistical perspective, those features also exhibit {intermediate to strong association} with the wind speed observations, {with Pearson's correlations} ranging on average between $0.87$ for wind gust and $-0.20$ for surface temperature. 

In addition to those six variables, we construct additional features that 
are not readily forecast by mesoscale models, but are derived from NWP outputs. We postulate the construction of two additional features: the spatio-temporal pressure differential and the geostrophic wind, which are described below.  

\subsubsection*{Spatio-temporal pressure differentials:}
Coarsely simplifying the physical processes involved, winds are generated by the movement of air from high- to low-pressure locations; the larger the pressure difference, the stronger the winds. 
If $P(\mathbf{s}_i, t)$ denotes the pressure at location $\mathbf{s}_i$ at time $t$, then, the spatio-temporal pressure differential between two locations $\mathbf{s}_i$ and $\mathbf{s}_j$ at two time instances $t$ and $t + d$, is defined as:
\begin{equation}
\label{lagPD}
\Delta_p(\mathbf{s}_i, \mathbf{s}_j,d)=P(\mathbf{s}_{i},t)-P(\mathbf{s}_{j},{t+d}),
\end{equation}
where $d \in \mathbb{Z}$ is a time lag. Note that $d$ can be both positive or negative, that is, we consider both past and future lags.  

\subsubsection*{Geostrophic wind:} \label{sec:gw}
The geostrophic wind is the flow that results from a balance between the Coriolis acceleration from the Earth’s rotation and the horizontal pressure gradient force at relatively high altitudes ($\sim$\hspace{.05mm}$1000$m). 
Geostrophic wind has been shown to improve the accuracy of short-term wind speed forecasts \cite{GWZhu}. Specifically, the geostrophic wind can be expressed as in (\ref{eqgw}), with $u_g$ and $v_g$ denoting its eastward and northward components, respectively.
\begin{equation}\label{eqgw}
\begin{gathered}
u_{g}=-\frac{g}{2 \Omega \sin \rho} \frac{\partial H}{\partial y_{c}}, \\
v_{g}=\frac{g}{2 \Omega \sin \rho} \frac{\partial H}{\partial x_{c}},
\end{gathered}
\end{equation}
where $H$ is the geopotential height, $g$ is the gravitational acceleration, $\Omega$ is the Earth rotation rate, $\rho$ is latitude, and $x_c$ and $y_c$ are local eastward and northward Cartesian coordinates. To compute $u_g$ and $v_g$, the geopotential height $H$ is first estimated through the hydrostatic equation, which is expressed for isothermal atmosphere as in (\ref{eqH}). 
\begin{equation}\label{eqH}
H=H_{i}+\frac{R \bar{T}}{g} \ln \left(\frac{p_{i}}{p_{0}}\right),
\end{equation}
where $H_i$, $p_i$ are the geopotential height and pressure at barometer $i$, respectively, $p_0$ is the reference pressure (850 hPa), $R$ is the gas constant (287 JK\textsuperscript{-1} kg\textsuperscript{-1}) and $\bar{T}$ is the layer-averaged temperature between $p_i$ and $p_0$, which we approximate by the surface temperature. {We acknowledge that relaxing the assumption of isothermal atmosphere by leveraging temperature information from RU-WRF may potentially provide more physically meaningful estimates of the geostrophic winds. This is an area of ongoing research.}

From there, we follow a similar procedure to that proposed in \cite{GWZhu} where a spatial response surface of the geopotential height, as in (\ref{eqss}), is estimated. 
A video showing the geopotential height response surface over time is included in the supplemental materials appended to this article (SM-1). 
\begin{equation} \label{eqss}
H\left(x_{c}, y_{c}\right)=c_{0}+c_{1} x_{c}+c_{2} y_{c}. 
\end{equation}

Setting $\frac{\partial H}{\partial x_{c}}=c_{1} \text { and } \frac{\partial H}{\partial y_{c}}=c_{2}$, we can compute $u_g$ and $v_g$, both of which are then used to estimate the final geostrophic wind as 
$Y_{g}(\mathbf{s}, t)=\sqrt{u_{g}^{2}+v_{g}^{2}}$. 

\subsection{Selecting the ``right'' features in $\mu(\mathbf{s}, t)$} \label{procedure} 
Putting the above pieces together, we have the following eight features as the potential constituents of $\tilde{\mathbf{G}}(\mathbf{s}, t)$: wind gust, air pressure, surface temperature, relative humidity, eastward and northward wind components, spatio-temporal pressure differential, and geostrophic wind. We also consider including lagged versions for each of those eight variables, as motivated by their potential role in correcting temporal biases. 

Suppose we only include four hourly lags for each of the eight variables listed above. This corresponds to $24$ lags in $10$-min resolution ($4$ hours $\times$ $6$ ten-minute intervals per hour). Hence, 
we end up with $8$ variables $\times$ $24$ lags = $192$ regressors for inclusion in $\tilde{\mathbf{G}}(\mathbf{s}, t)$. 
However, not all features are expected to be relevant at all times. In fact, more often than not, using an excessively large set of predictors does not coincide with the best predictive performance \cite{feng2017data} (the law of parsimony in ML). 
This also aligns with the {prior physical knowledge}: The drivers of NWP bias change over space-time, resulting in distinct NWP bias types and magnitudes. Thus, including a feature in the set $\tilde{\mathbf{G}}(\mathbf{s}, t)$ at a certain time does not justify its inclusion at other time instances. 
A dynamic feature selection mechanism is therefore needed to continuously identify and update a minimally sized subset of information-rich exogenous variables, constituting the elements of the set $\tilde{\mathbf{G}}(\mathbf{s}, t)$.

Given the time resolution at which the first set of forecasts are to be made ($10$-min ahead), advanced feature selection techniques that rely on iterative 
model estimation are practically prohibitive. Thus, we revert to simple (but effective) measures of explanatory power, namely partial autocorrelation functions (PACFs) to determine the time lag $\ell$ of $\tilde{\mathbf{Y}}^\ell$, and Pearson's correlation to select the features in $\tilde{\mathbf{G}}(\mathbf{s}, t)$. To ensure parsimony and avoid multicollinearity, we impose a simple rule: For the features in $\tilde{\mathbf{G}}(\mathbf{s}, t)$, we only select the most correlated lagged version of the same variable, i.e. the one that has the maximal correlation with the target response. 

\begin{figure}[h]
\centering  
\includegraphics[trim = 1cm 0.25cm 1cm 0.5cm, width =1 \linewidth]{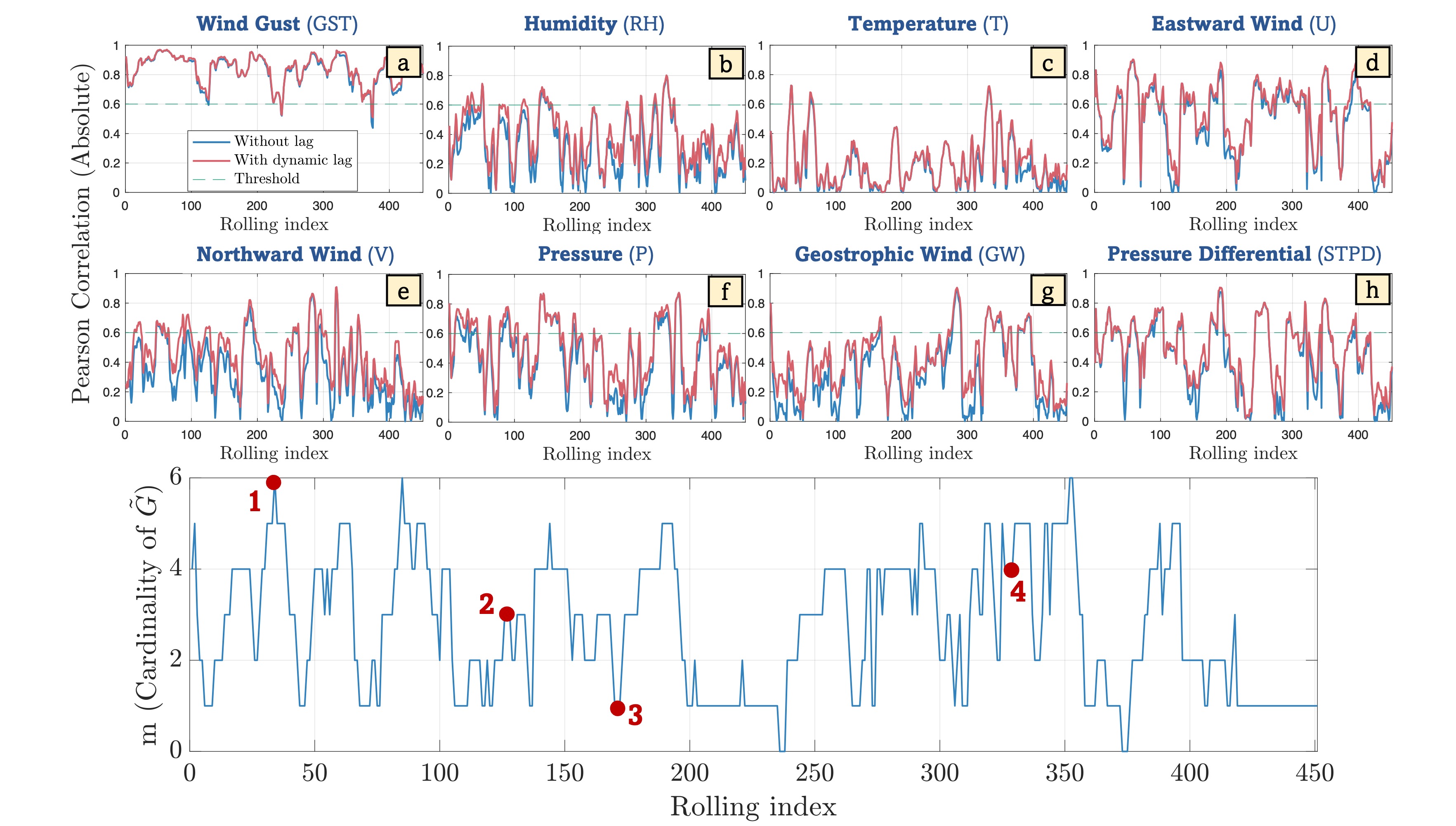}

\caption{(a)-(h) Pearson's correlations between eight variables and the wind speeds over $451$ forecast rolls (The green horizontal line depicts a threshold of $0.6$). Blue and red lines correspond to the correlation of the variable (with the predictand) versus that of a lagged version thereof. 
Panel (k) shows the change in $m$ (cardinality of $\tilde{\mathbf{G}}$) over time. Red points $1$-$4$ correspond to the following selected subsets (subscript denotes the chosen lag in $10$-min intervals): (1)  $\tilde{\mathbf{G}} =$ \{GST\textsubscript{-6}, RH\textsubscript{-18}, T\textsubscript{7}, U\textsubscript{10}, P\textsubscript{-18}, STPD\textsubscript{1}\}; (2) $\tilde{\mathbf{G}} =$ \{GST\textsubscript{-2}, U\textsubscript{-4}, STPD\textsubscript{-1}\}; (3): $\tilde{\mathbf{G}} = $\{GST\textsubscript{-3}\}; (4): $\tilde{\mathbf{G}} = $\{GW\textsubscript{0}, GST\textsubscript{10}, RH\textsubscript{-14}, P\textsubscript{18}\}}
\label{figcorprofile}  
\end{figure}

Figure \ref{figcorprofile}(a-h) shows the Pearson's correlations (in absolute value) between all eight variables of Section \ref{sec:mu} and the actual wind speed observations across all of the forecasting rolls. Looking at Figure \ref{figcorprofile}, two insights are immediately drawn. First, it is clear how the explanatory power of a feature can significantly vary over time. As an example, relative humidity (RH) can record correlations that reach up to $\sim 0.8$ at some time instances (Rolls 320-330), while having extremely low correlations at others (around Rolls 95-100). This suggests that this feature should only be included when it matters, i.e. at times when it can positively contribute towards explaining the variability in the predictand. Second, we note how selecting a lagged version of an exogenous feature can noticeably enhance its explanatory power. As a case in point, a lagged version of the northward wind component V (red solid line) appears to be a much more informative predictor (correlation around $0.6$, Roll $\sim$85) relative to the value of V itself (correlation around $0.45$; blue solid line). This is also valid with most features, in particular RH, V, P, GW, and STPD. Figure \ref{figcorprofile}(k) shows the change in $m$ (the number of features in $\tilde{\mathbf{G}}(\mathbf{s}, t)$) over time, showing how our feature selection dynamically adds or drops certain features over time, depending on their statistical relevance at the time of forecast, while keeping the input feature space reasonably sparse.
\subsection{Physics-guided modeling of  \(\eta(\mathbf{s}, t)\)}\label{sec:eta}
While $\mu(\mathbf{s}, t)$ is intended to capture large-scale fluctuations in the wind field, the role of $\eta(\mathbf{s}, t)$, on the other hand, is to characterize the higher-frequency variations (site-specific, minutes to hours) which are typically driven by sub-{mesoscale} local effects that NWPs may fail to capture. 
We decide to model $\eta(\mathbf{s}, t)$ as a spatio-temporal Gaussian Process (GP) 
\cite{cressie2015statistics, AzizGP}. Let $\mathbf{z}=\left[z\left(\mathbf{s}_{1}, t_{1}\right), z\left(\mathbf{s}_{1}, t_{2}\right), \ldots, z\left(\mathbf{s}_{1}, t_{T}\right), \ldots, z\left(\mathbf{s}_{n}, t_{T}\right)\right]^T$ be the vector of spatio-temporal residuals, such that $Z(\mathbf{s},t) = Y(\mathbf{s}, t) - \mu(\mathbf{s}, t)$. 
We regard the vector $\mathbf{z}$ as a realization of a spatio-temporal GP, $\mathcal{Z}(\cdot) \sim \mathcal{GP}(\mathcal{M}(\mathbf{s}, t), K({\boldsymbol{\gamma}},w))$, such that $\mathcal{M}(\mathbf{s}, t)$, and $K({\boldsymbol{\gamma}},w)$ are the GP mean and covariance (or kernel) functions, respectively, wherein 
${\boldsymbol{\gamma}} \in \mathbb{R}^2$ and $w \in \mathbb{Z}^+$ are the spatial and temporal lags, respectively,  while $\mathbb{Z}^+$ is the set of non-negative integers. 

The key challenge in GPs is to propose a suitable, mathematically permissible form for $K({\boldsymbol{\gamma}}, w): \mathbb{R}^2 \times \mathbb{Z}^+ \rightarrow \mathbb{R}$, which adequately captures the spatio-temporal dependence and enables GP-based forecasting. The most prevalent approach to specify $K({\boldsymbol{\gamma}},w)$ in the literature is through the so-called separable approach, which decomposes the dependence structure over space and time such that $K({\boldsymbol{\gamma}}, w) = K^{\mathbf{s}}({\boldsymbol{\gamma}}) \times K^t(w)$, wherein $K^{\mathbf{s}}({\boldsymbol{\gamma}})$ and $K^t(w)$ are two covariance structures for space and time, respectively \cite{cressie2015statistics}. Popular selections for $K^{\mathbf{s}}({\boldsymbol{\gamma}})$ and $K^t(w)$ include the squared exponential and Mat\'ern covariance functions \cite{bookGP}.

Despite its simplicity, the disconnect between space and time in the separable approach yields model specifications that violate the physical property of wind advection, i.e., the propagation of wind along a certain prevailing direction. This is because the separable approach assumes that space-time correlations are symmetric, i.e. $cor\big(Y(\mathbf{s}_i, t), Y(\mathbf{s}_{i'},t+\omega)\big) = cor\big(Y(\mathbf{s}_{i'},t), Y(\mathbf{s}_i, t + \omega)\big)$ \cite{SteinCov,AzizIEEE}. In reality, wind advection induces an asymmetry in the value of information, that is, along-wind dependence 
is expected to be stronger than opposite-wind dependence. In other words,  $cor\big(Y(\mathbf{s}_i, t), Y(\mathbf{s}_{i'},t+\omega)\big) - cor\big(Y(\mathbf{s}_{i'},t), Y(\mathbf{s}_i,t+\omega)\big) > 0$ when $i'$ is downstream of $i$. 

To capture this physical property, we propose to adopt a class of covariance models known in the geostatistical literature as the Lagrangian reference framework \cite{cox1988simple, salvana2022spatio}, which is capable of mimicking the ``advection'' of spatio-temporal information by having the following form: 
\begin{equation}\label{eq:lagr}
K({\boldsymbol{\gamma}},w)=\mathbb{E}_{{\mathrm{\Theta}}}\{\psi({\boldsymbol{\gamma}}-{{\mathbf{\Theta}}} w)\},
\end{equation}
where $\psi(\cdot)$ is a positive-definite kernel. The advection vector ${{\mathbf{\Theta}}} \in \mathbb{R}^2$ represents the prevailing flow and hence, its specification is instrumental to Lagrangian models. 
Here, we model {${\pmb{\Theta}}$} as a bivariate normal random vector, since wind is not only governed by advection, but also diffusion, i.e. the random propagation of wind in other directions \cite{SchlatherCov,salvana2020nonstationary}. Letting ${{\mathbf{\Theta}}}\sim\mathcal{N}_{d}\left({\pmb{\mu}_{{\Theta}}},{\pmb{\Sigma}_{\Theta}}\right)$
, $\psi(x)= \exp(-x^2)$, and $\mathbf{F} = \mathbf{I}_{d \times d}+2{\pmb{\Sigma}_{{\Theta}}} w^{2}$, yields the following expression for the Lagrangian correlation model: 
\begin{equation}\label{eq:lagr2}
K_{LG}({\boldsymbol{\gamma}},w)=\frac{1}{\sqrt{\left|
\mathbf{F}\right|}} \exp \left\{-\left({\boldsymbol{\gamma}}-{\pmb{\mu}_{{\Theta}}} w\right)^{T}
\mathbf{F}
^{-1}\left({\boldsymbol{\gamma}}-{\pmb{\mu}_{{\Theta}}} w\right)\right\},
\end{equation}
where $\left|\cdot\right|$ denotes the matrix determinant. 

We propose to estimate ${\pmb{\mu}_{{\Theta}}}$ and ${\pmb{\Sigma}_{{\Theta}}}$ in light of the NWP forecasts of the eastward and northward wind components as reasonable representations of the prevailing flow during the forecast horizon. As shown in (\ref{eq:nwp1}) and (\ref{eq:nwp2}), our estimates for ${\pmb{\mu}_{{\Theta}}}$ and ${\pmb{\Sigma}_{{\Theta}}}$ 
will be the spatio-temporal empirical averages and covariances of the NWP forecasts in the window $[t_c - T, t_c + {H_T}]$, where $t_c$ is the current time, while $T$ and ${H_T}$ are the lengths of the training data and of the forecast horizon, both in $10$-min intervals.  
\begin{equation}\label{eq:nwp1}
    {\pmb{\mu}_{{\Theta}}} = [\mu_1, \mu_2]^T = [\bar{{u}}, \bar{{v}}]^T, 
\end{equation}
\begin{equation}\label{eq:nwp2}
    {\pmb{\Sigma}_{{\Theta}}} =
\left[\begin{array}{cc}
\sigma_{1,1}   &   \sigma_{1,2} \\ \sigma_{2,1}   &   \sigma_{2,2}\end{array} \right] = \left[\begin{array}{cc}
cov(\mathbf{{u}}, \mathbf{{u}}) & cov(\mathbf{{u}}, \mathbf{{v}}) \\ cov(\mathbf{{v}}, \mathbf{{u}})   &   cov(\mathbf{{v}}, \mathbf{{v}})\end{array} \right],
\end{equation}
where $\mathbf{{u}}= [{u}_{t_c-T}, ..., {u}_{t_c+h}]^T$ and $\mathbf{v} = [v_{t_c - T}, ..., v_{t_c+h}]^T$ are the NWP outputs for the eastward and northward winds, respectively, during both the training and forecast horizon windows, whereas $\bar{{u}}$ and $\bar{{v}}$ are the sample means of $\mathbf{{u}}$ and $\mathbf{{v}}$, respectively, and $cov(\cdot, \cdot)$ denotes the sample covariance. 

Our final covariance function $K_{\eta}(\mathbf{s}, t)$ is a convex combination of a separable correlation function and the Lagrangian correlation function of (\ref{eq:lagr2}): 
\begin{equation}\label{eq:cov}
\small
K_{\eta}({\boldsymbol{\gamma}}, w)=
\alpha\bigg[\lambda \hspace{-0.4cm} \overbrace{\left(K_{SE}^{s} ({\boldsymbol{\gamma}})\times 
K_{SE}^t(w)\right)}^{
\parbox{11em}{\scriptsize \centering Separable Kernel}} \hspace{-0.35cm}+ 
(1-\lambda)\hspace{-1.2cm}\overbrace{K_{LG}({\boldsymbol{\gamma}}, w)}^{\parbox{11em}{\scriptsize \centering Lagrangian Kernel}}\hspace{-1.1cm}\bigg] 
+\mathbb{I}_{\{||{\boldsymbol{\gamma}}||=|w|=0\}} \delta,
\end{equation}
where $\alpha > 0$ is the marginal variance parameter, $\lambda \in [0,1]^T$ is the asymmetry coefficient denoting the strength of symmetry (or lack thereof), and $\delta > 0$ denotes the noise variance, whereas $\mathbb{I}(\cdot)$ is the indicator function and $||\cdot||$ is the Euclidean norm. The terms $K_{SE}^{\mathbf{s}}(\cdot)$ and $K_{SE}^t(\cdot)$ are squared exponential (SE) correlation functions for space and time, respectively. 

\subsection{Estimation and probabilistic forecasting}
The parameters to be estimated are: (1) $\mathbf{a}$, $\mathbf{b}$, and $\mathbf{c}$ denote the sets of parameters in the calibration term $\mu(\mathbf{s}, t)$ in (\ref{mu}); (2) ${\pmb{\mu}_{{\Theta}}}$ and ${\pmb{\Sigma}_{{\Theta}}}$ are the advection vector parameters; (3) {t}he marginal variance $\alpha$ and asymmetry parameter $\lambda$ in (\ref{eq:cov}), and the GP mean parameter $\mathcal{M}(\mathbf{s}, t) = \beta_0$; (4) {t}he range parameters of $K_{{SE}}^{\mathbf{s}}(\cdot)$ and $K_{SE}^t(\cdot)$, 
denoted by $r_{\mathbf{s}}$ and $r_t$, respectively; and (5) {t}he noise variance, $\delta$. We estimate the parameters sequentially: First, we estimate $\mathbf{a}$, $\mathbf{b}$, and $\mathbf{c}$ via least squares, then, use the residuals to estimate the remainder of the parameters using maximum likelihood estimation, except for ${\pmb{\mu}_{{\Theta}}}$ and ${\pmb{\Sigma}_{{\Theta}}}$ which are estimated as in (\ref{eq:nwp1})-(\ref{eq:nwp2}). 

The joint predictive distribution of the final set of spatio-temporal forecasts, $\hat{\mathbf{f}} = [\hat{f}(\mathbf{s}_1, t_c + 1), ..., \hat{f}(\mathbf{s}_n, t_c+{H_T})]^T$, is fully characterized by virtue of the GP framework. 
In specific, the predictive mean and variance at each look-ahead time and spatial location $\hat{f}(\mathbf{s}_i, t_c + h)$ are obtained as: 
\begin{equation}\label{eq:gpmean}
    \hat{f}(\mathbf{s}_i, t_c + h)  = 
    \hat{\mu}(\mathbf{s}_i, t_c+h) + \hat{\beta}_0 + 
    \hat{\mathbf{k}}^T 
    \bar{\mathbf{K}}_{\eta}^{-1}
    (\mathbf{y} - \bar{\pmb{\mu}} - \hat{\beta}_0\mathbf{I}), 
\end{equation}
\begin{equation}\label{eq:gpvar}
\begin{split}
\hat{\sigma}^2(\mathbf{s}_i, t_c+h)  = &
\hat{K}_{\eta}(\mathbf{0},0) -  \hat{\mathbf{k}}^T 
\bar{\mathbf{K}}_{\eta}^{-1}
\hat{\mathbf{k}}  \\ &   + (1 - (\hat{\mathbf{k}}^T 
\bar{\mathbf{K}}_{\eta}^{-1}
\mathbf{I}))^T(\mathbf{I}^T 
\bar{\mathbf{K}}_{\eta}^{-1}
\mathbf{I})^{-1}(1 - [\hat{\mathbf{k}}^T 
\bar{\mathbf{K}}_{\eta}^{-1}
\mathbf{I}]),
\end{split}
\end{equation}
where $h \in \{1, ..., {H_T}\}$ is the forecast horizon (in $10$-min intervals), $\mathbf{I}$ is an $n \cdot {H_T} \times 1$ column of $1$'s, and $\bar{\pmb{\mu}} = [\mu(\mathbf{s}_1, t_1), ..., \mu(\mathbf{s}_n, t_c)]^T$ is the vector of evaluations of $\mu(\mathbf{s}, t)$ for the training data. Similarly, $\bar{\mathbf{K}}_{\eta}$ is the training covariance matrix evaluated using the estimated kernel $\hat{K}_{\eta}(\cdot, \cdot)$. The vector $\hat{\mathbf{k}}$ contains the pairwise covariances between $\mathbf{z} = \mathbf{y} - \bar{\pmb{\mu}}$ and $z(\mathbf{s}_i, t_c + h)$. 

Two unique advantages of AIRU-WRF are: (1) its ability to make full probabilistic inference about the joint distribution of the spatio-temporal forecasts \cite{arrieta2022spatio}, which can be used to generate trajectories that naturally embed the spatial and temporal dependence for operational decisions (e.g., for use within a stochastic program) \cite{Pinson2013, petros2022}
; (2) The ability to produce ``spatial wind field forecast maps,'' in the form of two-dimensional images, including at locations where no measurements are available. 
Both of those capabilities are demonstrated in Section 4. 

\section{Real-world Experiments and Discussions}\label{results}
We focus on short-term forecasts (up to six hours) in $10$-min resolution, i.e. $h \in \{1, ..., 36\}$. We evaluate AIRU-WRF at E05 and E06 (where data are available), and visualize its forecasts at nearby sites where no data exist. 

\subsection{Training and testing} 
We test our approach using a rolling forecasting scheme, wherein for each forecast roll, we perform our feature selection procedure, re-train the model, obtain the forecasts, then roll by $6$ hours, and repeat the whole process. 
For the four-month winter period, this is correspondent to $451$ rolls, yielding a total of $6$ forecasts/hour $\times$ $6$-hour horizon $\times$ $451$ rolls $\times$ $2$ spatial sites = $32,472$ testing instances. 
For the two-month summer period, this is correspondent to $216$ rolls, yielding a total of $6$ forecasts/hour × $6$-hour horizon × $216$ rolls × $2$ spatial sites = $15,552$ testing instances. 
We find that five days of historical data is a sufficient training data size to balance model fitting and computational efficiency. 
The maximum time to complete one forecasting roll (including feature selection, training, and forecasting) was $2.52$ minutes on a standard laptop with an \texttt{i7-6700HQ} Intel processor, $2.60$GHz base frequency, and $16$ GB RAM. We also find that using NWPs for sub-hourly forecasts (i.e., $h < 6$, or ultra-short-term forecasts) is, on average, not helpful. So we exclude the term $\mu(\mathbf{s}, t)$ when making sub-hourly forecasts, and include it for all other forecast horizons (i.e. for $h \geq 6$).

\subsection{Benchmarks and evaluation metrics}
We compare AIRU-WRF against five benchmarks, $\mathcal{B}1$ - $\mathcal{B}5$: 

\begin{itemize}
\item [($\mathcal{B}1$)] GOP: The Geostatistical Output Perturbation (GOP), proposed in \cite{Gel}, is a hybrid model to calibrate {mesoscale} NWPs using local observations. GOP is regarded as a special case of AIRU-WRF by assuming $\mu^{\text{GOP}}(\mathbf{s}, t)=\mathbf{a}^T \tilde{\mathbf{H}}(\mathbf{s}, t)+ (\mathbf{b}^T \tilde{\mathbf{H}}(\mathbf{s}, t))\tilde{{Y}}(\mathbf{s},t)$, where $\tilde{\mathbf{H}}(\mathbf{s}, t)$ is the set of NWP variables directly obtained from RU-WRF (namely, air pressure, surface temperature, wind gust, relative humidity, eastward and northward wind components), while  $\eta^{GOP}(\mathbf{s},t)$ is assumed to be a GP with a spatial SE kernel. AIRU-WRF generalizes on GOP on two main fronts: (1) Extension to the spatio-temporal setting via a physically meaningful covariance function; (2) A physics-guided calibration of the NWPs through the construction and dynamic updating and selection of physically meaningful predictors encoded in the set $\tilde{\mathbf{G}}(\mathbf{s}, t)$. 
\item[($\mathcal{B}2$)] NWP: Those are the raw forecasts from RU-WRF (statistically interpolated to the $10$-min resolution). 

\item[($\mathcal{B}3$)] ARIMAX($p$,$q$,$d$): We train a separate Autoregressive Integrated Moving Average with Exogenous inputs (ARIMAX) model for each forecast location, with $\tilde{\mathbf{H}}$ as the set of exogenous inputs (the same set used in GOP). All model parameters ($p$, $q$, and $d$) as well as the autoregressive, moving average, and exogenous coefficients 
are optimized using the \texttt{pmdarima} package in \texttt{Python}. 
\item[($\mathcal{B}4$)] LSTM: Long short-term memory networks is a class of deep learning models based on recurrent neural networks that is well-suited for time series data \cite{ko2020deep}. We fit a separate LSTM for each location  and use  stochastic gradient descent to optimize the LSTM network parameters through the \texttt{Deep Learning Toolbox} in \texttt{Matlab}. A grid search was used to tune the number of hidden units for the LSTM (set at $100$), number of epochs (set at $100$), and the learning rate (start = $0.005$, decreased by a factor of $0.1$ every $30$ epochs).  

\item[($\mathcal{B}5$)] PER: Persistence (PER) forecasting is a standard benchmark that assumes present weather conditions will persist into the future. 
It is known to perform well at ultra-short-term horizons. 

\end{itemize}

To evaluate the forecasts, we use 
the Mean Absolute Error (MAE) 
and the Continuous Ranked Probability Score (CRPS) for point and probabilistic wind speed forecasting, respectively. 
We only compute CRPS for probabilistic approaches, namely AIRU-WRF, GOP, and ARIMAX. {We convert the wind speed forecasts from all competing models into wind power predictions using statistically constructed wind power curves (more on that in Section 4.4).} We use the power curve error (PCE) loss to evaluate the wind power predictions \cite{Hering}.

\subsection{Wind speed forecasting results} 
Tables \ref{tab:MAE} and \ref{tab:MAE_2} show the MAE and CRPS values for the six models (including AIRU-WRF), at both sites, aggregated in hourly intervals for winter and summer periods, respectively. Overall, AIRU-WRF outperforms statistical methods (ARIMAX, PER) by $16.8$-$27.1$\%, physics-based models (NWP) by $16.6$-$22.5$\%, hybrid methods (GOP) by $18.5$-$21.4$\%, and deep learning-based methods (LSTM) by $19.1$-$29.1$\%. A deeper analysis of the results in Tables \ref{tab:MAE} and \ref{tab:MAE_2} yields useful insights, as discussed next.
  
\begin{table*}[]
    \centering
\caption{\centering Wind speed forecasting results for the four-month winter period (November, 2019 to February, 2020): MAE (left) and CRPS (right), aggregated in hourly intervals, for sites E05 (Top) and E06 (Bottom). Bold-faced values denote the best performance.} 
\resizebox{\columnwidth}{!}{%
\begin{tabular}{|c| c c c c c c|| c c c|}\hline 
            & \multicolumn{9}{c|}{\textbf{E05 (39°58'10"N and 72°43'00"W)}} \\
            \hline
            & \multicolumn{6}{c||}{\textbf{MAE}}   & \multicolumn{3}{c|}{\textbf{CRPS}}   \\
            \hline
            Horizon (hrs) & AIRU-WRF &  GOP &	NWP & ARIMAX  & LSTM & PER    & AIRU-WRF & GOP & ARIMAX \\\hline
            1 & 0.753           & $0.922$  &$1.657$ & $0.794$ & $0.791$ & $\mathbf{0.743}$ & $\mathbf{0.575}$  &   $0.742$  &   $0.643$  \\
            2 &	$\mathbf{1.267}$ & $1.628$  & $1.601$ &	$1.333$ & $1.300$  & $1.287$ &   $\mathbf{0.957}$    & $1.205$   &   $1.039$   \\
            3 &	$\mathbf{1.451}$ & $1.750$  & $1.592$ &	$1.587$ &	$1.716$  & $1.708$  &   $\mathbf{1.094}$ &   $1.282$ &   $1.212$\\
            4 &	$\mathbf{1.478}$ & $1.798$  & $1.586$ &	$1.860$ &	$2.150$  & $2.164$  & $\mathbf{1.133}$  & $1.319$  &  $1.407$\\
            5 &	$\mathbf{1.561}$ & $1.859$  & $1.591$ &	$2.055$ &	$2.499$  & $2.495$  & $\mathbf{1.186}$  & $1.358$  & $1.536$ \\
            6 &	$\mathbf{1.651}$ & $2.052$  & $1.741$ &	$2.287$ &	$2.782$ & $2.801$ & $\mathbf{1.274}$  &  $1.490$  &  $1.656$\\
      Average & $\mathbf{1.360}$ & $1.668$  &	$1.631$ & $1.653$ &   $1.873$ & $1.866$ & $\mathbf{1.037}$  & $1.233$  & $1.249$\\
            \% Improvement & - & $18.5$\%  & $16.6$\% & $17.7$\% &  $27.4$\% & $27.1$\% &  - & $15.9$\%  & $17.0$\% \\
        \Xhline{1\arrayrulewidth}
    \quad
   
           & \multicolumn{9}{c|}{\textbf{E06 (39°32'50"N and 73°25'45"W)}} \\
            \hline 
                        & \multicolumn{6}{c||}{\textbf{MAE}}   & \multicolumn{3}{c|}{\textbf{CRPS}}   \\
            \hline
            Horizon (hrs) & AIRU-WRF & GOP &	NWP & ARIMAX  & LSTM & PER    &   AIRU-WRF  &   GOP & ARIMAX    \\\hline
            1 & $\mathbf{0.727}$ & $0.975$  &	$1.621$ &   $0.767$ & $0.805$ & $0.729$ & $\mathbf{0.556}$  & $0.753$  &  $0.614$\\
            2 &	$\mathbf{1.271}$ & $1.702$  &   $1.691$ &	$1.347$ & $1.372$  & $1.277$ & $\mathbf{0.969}$  & $1.291$  &  $1.018$\\ 
            3 &	$\mathbf{1.530}$ & $1.902$  &	$1.730$ &	$1.663$ & $1.855$  & $1.753$ & $\mathbf{1.193}$  & $1.447$  &  $1.256$\\
            4 &	$\mathbf{1.558}$ & $1.978$  &	$1.801$ &	$1.942$ & $2.235$  & $2.156$ & $\mathbf{1.216}$  & $1.470$  &  $1.463$\\
            5 &	$\mathbf{1.592}$ & $1.973$  &	$1.706$ &	$2.077$ & $2.566$  & $2.504$ & $\mathbf{1.235}$  & $1.440$  &  $1.515$\\
            6 &	$\mathbf{1.584}$ & $1.988$  &	$1.659$ &	$2.137$ & $2.827$  & $2.779$ & $\mathbf{1.252}$  & $1.441$  &  $1.557$\\
      Average & $\mathbf{1.377}$ & $1.753$  &	 $1.701$  &   $1.656$ & $1.943$ & $1.866$ &  $\mathbf{1.070}$  & $1.307$  &  $1.237$\\
            \% Improvement & - &	$21.4$\%  &	$19.1$\% & $16.8$\% &  $29.1$\% & $26.2$\% & -  & $18.1$\% & $13.5$\% \\ \hline
        \end{tabular}
        }
    \label{tab:MAE}
\end{table*}

\begin{table*}[]
    \centering
\caption{\centering Wind speed forecasting results for the two-month summer period (May, 2020 to June, 2020): MAE (left) and CRPS (right) values, aggregated in hourly intervals, for sites E05 (Top) and E06 (Bottom). Bold-faced values denote the best performance.} 
\resizebox{\columnwidth}{!}{%
\begin{tabular}{|c| c c c c c c|| c c c|}\hline 
            & \multicolumn{9}{c|}{\textbf{E05 (39°58'10"N and 72°43'00"W)}} \\
            \hline
            & \multicolumn{6}{c||}{\textbf{MAE}}   & \multicolumn{3}{c|}{\textbf{CRPS}}   \\
            \hline
            Horizon (hrs) & AIRU-WRF &  GOP &	NWP & ARIMAX  & LSTM & PER    & AIRU-WRF & GOP & ARIMAX \\\hline
            1 & $\mathbf{0.763}$ & $1.176$  & $2.086$ & $0.904$ &   $0.895$  & $0.817$  & $\mathbf{0.614}$  & $0.931$  &   $0.729$  \\
            2 &	$\mathbf{1.413}$ & $2.044$  & $2.120$ &	$1.649$ &   $1.558$  & $1.471$  & $\mathbf{1.052}$  & $1.548$  &   $1.319$   \\
            3 &	$\mathbf{1.702}$ & $2.054$  & $2.048$ &	$1.999$ &	$1.988$  & $1.849$  & $\mathbf{1.293}$  & $1.484$  &   $1.545$\\
            4 &	$\mathbf{1.872}$ & $2.193$  & $2.095$ &	$2.295$ &	$2.338$  & $2.206$  & $\mathbf{1.501}$  & $1.518$  &  $1.688$\\
            5 &	$\mathbf{1.987}$ & $2.301$  & $2.158$ &	$2.461$ &	$2.532$  & $2.490$  & $\mathbf{1.565}$  & $1.731$  & $1.703$ \\
            6 &	$\mathbf{1.997}$ & $2.441$  & $2.023$ &	$2.621$ &	$2.722$ & $2.760$   & $\mathbf{1.566}$  & $1.851$  &  $1.794$\\
      Average & $\mathbf{1.622}$ & $2.035$  &	$2.088$ & $1.988$ &   $2.006$ & $1.932$   & $\mathbf{1.265}$  & $1.511$  & $1.463$\\
            \% Improvement & - & $20.3$\%  & $22.3$\% & $18.4$\% &  $19.1$\% & $16.0$\% &  - & $16.2$\%  & $13.5$\% \\
        \Xhline{1\arrayrulewidth}
    \quad
   
           & \multicolumn{9}{c|}{\textbf{E06 (39°32'50"N and 73°25'45"W)}} \\
            \hline 
                        & \multicolumn{6}{c||}{\textbf{MAE}}   & \multicolumn{3}{c|}{\textbf{CRPS}}   \\
            \hline
            Horizon (hrs) & AIRU-WRF & GOP &	NWP & ARIMAX  & LSTM & PER    &   AIRU-WRF  &   GOP & ARIMAX    \\\hline
            1 & $\mathbf{0.787}$ & $1.103$  &	$2.098$ &   $0.856$ & $0.882$  & $0.805$ & $\mathbf{0.636}$  & $0.913$  &  $0.685$\\
            2 &	$\mathbf{1.501}$ & $1.856$  &   $2.059$ &	$1.671$ & $1.593$  & $1.514$ & $\mathbf{1.072}$  & $1.408$  &  $1.267$\\ 
            3 &	$\mathbf{1.807}$ & $2.085$  &	$2.115$ &	$2.030$ & $2.056$  & $1.958$ & $\mathbf{1.382}$  & $1.530$  &  $1.570$\\
            4 &	$\mathbf{1.870}$ & $2.251$  &	$2.115$ &	$2.228$ & $2.361$  & $2.266$ & $\mathbf{1.478}$  & $1.583$  &  $1.629$\\
            5 &	$\mathbf{1.880}$ & $2.342$  &	$2.085$ &	$2.380$ & $2.612$  & $2.505$ & $\mathbf{1.495}$  & $1.702$  &  $1.706$\\
            6 &	$\mathbf{1.951}$ & $2.583$  &	$2.172$ &	$2.618$ & $2.901$  & $2.809$ & $\mathbf{1.580}$  & $1.956$  &  $1.815$\\
      Average & $\mathbf{1.633}$ & $2.037$  &	  $2.107$ &   $1.964$ & $2.068$ & $1.976$ &  $\mathbf{1.274}$  & $1.515$  &  $1.445$\\
            \% Improvement & - &	$19.8$\%  &	$22.5$\% & $16.9$\% &  $21.0$\% & $17.4$\% & -  & $15.9$\% & $11.9$\% \\ \hline
        \end{tabular}
        }
    \label{tab:MAE_2}
\end{table*}

\begin{figure}  
\centering  
\includegraphics[trim = 1cm 0.25cm 1cm  0.5cm, width = 1 \linewidth]{MAEs5.jpg}
\caption{Average MAE (across $451+216=667$ forecast rolls), aggregated in hourly intervals, for the forecasts made via AIRU-WRF (blue circles), NWP (red triangles), and ARIMAX (green squares) for E05 (left) and E06 (right). \vspace{-0.35cm}}
\label{figMAE}  
\end{figure}  

First, statistical methods (ARIMAX, PER) appear to perform well in ultra-short-term horizons (first three hours, $h < 18$) relative to NWPs. {This aligns with the general consensus in the wind forecasting literature \cite{sweeney2020future, ye2023b}}. The trend is reversed for longer forecast horizons (last three hours, $h > 18$) by virtue of the embedded physics within NWPs enabling them to reliably extrapolate at longer horizons. AIRU-WRF, on the other hand, outperforms both sets of methods in almost all forecast horizons (with the exception of the first hour in E05, where AIRU-WRF comes in as a close second to PER). This is clearly demonstrated in Figure \ref{figMAE} showing how AIRU-WRF borrows strength across the physical and statistical learning paradigms, yielding forecasts that are superior to both across all forecast horizons. 

Second, Figure \ref{figMAE}, as well as Tables \ref{tab:MAE} and \ref{tab:MAE_2} show that the improvement from AIRU-WRF over NWPs is maximal in short-term horizons where the ``learning from data'' aspect of AIRU-WRF provides a significant advantage relative to a purely physics-based model. This margin of improvement decays as the forecast horizon becomes longer, with forecasts from AIRU-WRF gradually converging towards their NWP counterparts, yet still maintaining a non-negligible lead, especially for E06. 
On the contrary, the advantage of AIRU-WRF over purely statistical methods peaks at longer forecast horizons (up to $43$\% at the sixth hour), where the embedded physics in AIRU-WRF enable it to safely extrapolate at longer horizons where solely depending on data-driven learning is typically unreliable. 

Third, comparing a deep learning (DL) method like LSTM with statistical methods (ARIMAX, PER) 
suggests that their performance is comparable in short-term forecast horizons ($\sim$\hspace{-0.02mm} first four hours), with the latter gradually overtaking the former as the forecast horizon extends. We speculate that DL may need a much larger dataset (in terms of the number of locations and measurements) to unleash its full potential. Having a dense network of observations in the U.S. {Mid} Atlantic (and other similarly under-explored OSW areas) is, however, not practical, at least in the foreseeable future. AIRU-WRF, on the other hand, performs significantly better than all physics-free methods (be it statistical- or DL-based), in almost all forecast horizons.

Finally, comparing AIRU-WRF to a seminal hybrid method (GOP) demonstrates the merit of the physics-guided modeling of  $\mu(\mathbf{s}, t)$ and $\eta(\mathbf{s}, t)$. The largest improvement from AIRU-WRF relative to GOP is realized at shorter forecast horizons, where the benefit of invoking a physically meaningful spatio-temporal kernel materializes. Noticeable improvements at longer horizons are still maintained, and are likely attributed to the construction and integration of physically relevant features within AIRU-WRF. 

\begin{figure}  
\centering  
\includegraphics[trim = 1cm .25cm 1cm 5cm, width=0.95 \linewidth]{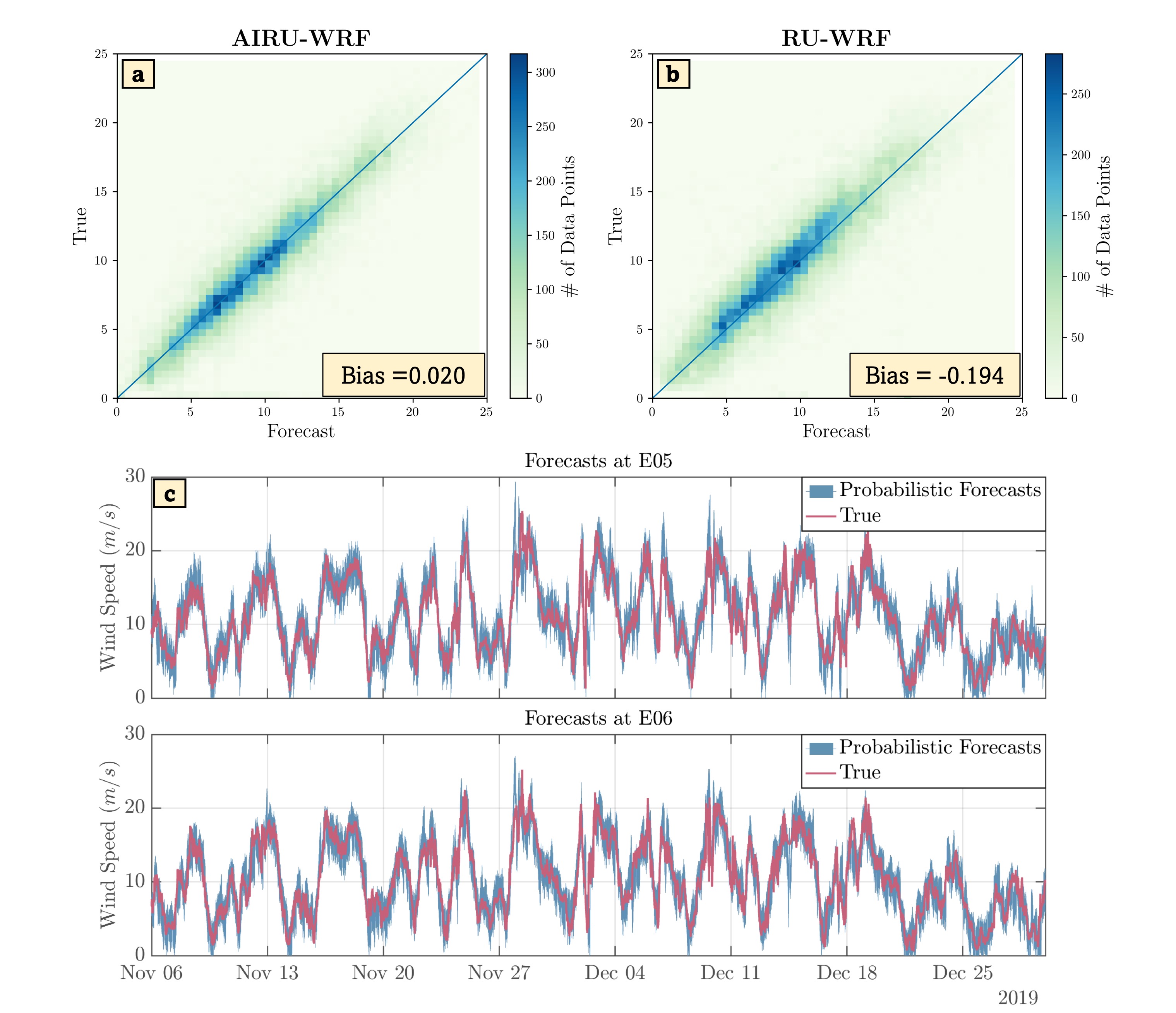}
\caption{(a)-(b) Comparing AIRU-WRF and RU-WRF: On average, 
AIRU-WRF's forecasts are noticeably closer to the true values, yielding a $\sim$ \hspace{-.02cm}$90$\% bias reduction (defined as the average signed difference between data and forecasts). (c) AIRU-WRF's probabilistic forecasts (10\% and 90\% percentiles shown), on top of the true observations.} 
\label{fig:forecast_all}  
\end{figure}

\begin{figure}  
\centering  
\includegraphics[trim = .5cm .5cm 0cm .5cm, width=.95\linewidth]{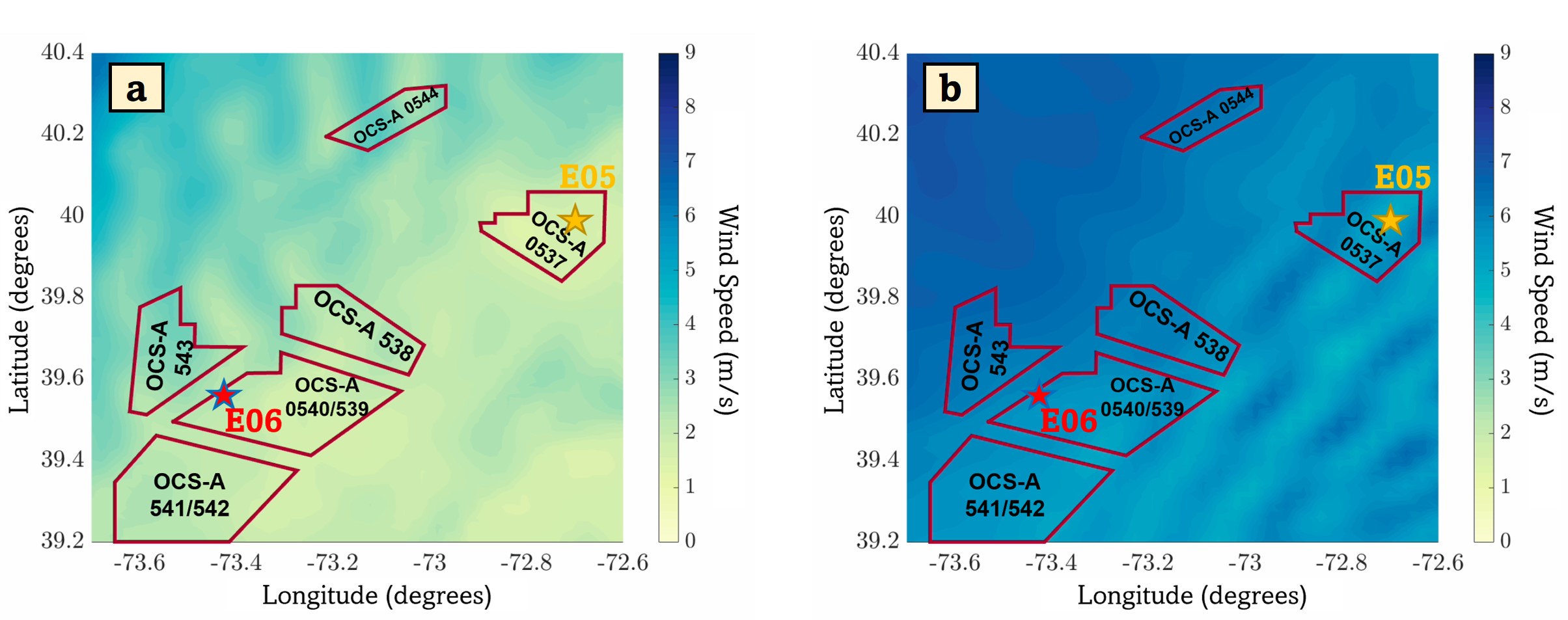}
\caption{AIRU-WRF's spatial wind field forecast maps as evolving $2$-dimensional images at two separate time instances on the same day. Stars denote E05 and E06 locations, while red polygons roughly depict the OSW energy lease areas, as of August, 2021 \cite{lease2021}.  }
\label{figfield}  
\end{figure}

Figure \ref{fig:forecast_all}(c) shows AIRU-WRF's probabilistic forecasts suggesting a faithful alignment with actual observations. The ability of AIRU-WRF to adjust forecast biases is demonstrated in Figure \ref{fig:forecast_all}(a)-(b), where the true versus forecast values for AIRU-WRF 
show a more symmetric clustering around the $45^\circ$ line, relative to those from RU-WRF. 
Moreover, unique to AIRU-WRF is its ability to make forecasts at locations where no observations are available. Those are used to produce spatial wind field forecast ``maps'', in the form of evolving two-dimensional images for a region of interest. Figure \ref{figfield} shows examples of those wind field forecast maps at two separate time instances on a select day, on top of the OSW energy lease areas in the NY/NJ Bight. A video showing the evolution of those wind field forecast maps is included in the supplemental materials (SM-2). Those forecast maps can be highly effective in communicating AIRU-WRF's outputs to key stakeholders in the OSW energy industry.

\subsection{Wind power forecasting results}

In order to showcase the value of our approach to the OSW energy industry, we convert the wind speed forecasts obtained via AIRU-WRF into wind power predictions. As there are no wind farms currently located in the NY/NJ Bight region, we make use of actual power curves constructed via the method of bins \cite{international2017wind, golparvar2021surrogate} using SCADA data from an operational wind farm in the United States \cite{ding2019data}. The power output is then scaled to the $[0,1]$ interval, where the maximum rated capacity is represented by a value of $1$. Using the constructed power curve, we transform the corresponding wind speed forecasts from the six competing models into wind power predictions. 

The resulting power predictions are evaluated using the PCE loss \cite{Hering}, which assigns unequal weights for under- and over-prediction, as in (\ref{eq:pce}).
\begin{equation}\label{eq:pce}
P C E(P, \widehat{P})=\left\{\begin{array}{c}
g\left[P(\mathbf{s},t_c+h)-\widehat{P}(\mathbf{s},t_c+h)\right] \\
\text { if } \hat{f}(\mathbf{s}, t_c + h) \leq Y(\mathbf{s},t_c+h), \\
(1-g)\left[\widehat{P}(\mathbf{s},t_c+h)-P(\mathbf{s},t_c+h)\right] \\
\text { if } \hat{f}(\mathbf{s}, t_c + h) >Y(\mathbf{s},t_c+h),
\end{array}\right.
\end{equation}
where $P(\mathbf{s},t_c+h)$ and $\widehat{P}(\mathbf{s},t_c+h)$ are the normalized power observations and forecasts at $t_c+h$ and the $\mathbf{s}$th location, and $g$ is the under-estimation weight, 
which is typically set at values higher than $0.5$. 
Table \ref{tab:pce_power}, shows the average PCE values across all horizons for values of $g$ ranging between $0.5$ and $0.8$ with $0.1$ increment, as well as $g$ = 0.73 (the value reported in \cite{Pinson2007, Hering}), for sites E05 and E06. AIRU-WRF is shown to significantly outperform all of its competitors, suggesting that the improvements attained in wind speed forecasting are translated into predictive gains in the wind power domain. 

\begin{table*}[]
    \centering
\caption{\centering Wind power forecasting results for the combined winter and summer periods (total of six months): Average PCE values for sites E05 (Top) and E06 (Bottom). Bold-faced values denote the best performance.} 
    \small
\begin{tabular}{|c| c c c c c c|}\hline 
            & \multicolumn{6}{c|}{\textbf{E05 (39°58'10"N and 72°43'00"W)}} \\
            \hline
            $g$ & AIRU-WRF &  GOP &	NWP & ARIMAX  & LSTM & PER     \\\hline
            0.5 & $\mathbf{0.051}$ & $0.061$  & $0.062$ & $0.063$ &   $0.071$  & $0.068$  \\
            0.6 &	$\mathbf{0.052}$ & $0.062$  & $0.065$ &	$0.064$ &   $0.070$  & $0.068$  \\
            0.7 &	$\mathbf{0.053}$ & $0.064$  & $0.069$ &	$0.064$ &	$0.070$  & $0.069$  \\
            0.73 &	$\mathbf{0.054}$ & $0.064$  & $0.070$ &	$0.065$ &	$0.070$  & $0.069$  \\
            0.8 &	$\mathbf{0.054}$ & $0.065$  & $0.071$ &	$0.065$ &	$0.070$  & $0.069$  \\
        \Xhline{1\arrayrulewidth}
    \quad
   
            & \multicolumn{6}{c|}{\textbf{E06 (39°32'50"N and 73°25'45"W)}}  \\
            \hline
             $g$ & AIRU-WRF &  GOP &	NWP & ARIMAX  & LSTM & PER     \\\hline
            0.5 & $\mathbf{0.052}$ & $0.064$  & $0.065$ & $0.066$ &   $0.076$  & $0.071$  \\
            0.6 &	$\mathbf{0.053}$ & $0.065$  & $0.067$ &	$0.066$ &   $0.076$  & $0.072$  \\
            0.7 &	$\mathbf{0.054}$ & $0.066$  & $0.069$ &	$0.066$ &	$0.075$  & $0.072$  \\
            0.73 &	$\mathbf{0.054}$ & $0.066$  & $0.070$ &	$0.067$ &	$0.075$  & $0.072$  \\
            0.8 &	$\mathbf{0.054}$ & $0.067$  & $0.071$ &	$0.067$ &	$0.075$  & $0.073$  \\
            \Xhline{1\arrayrulewidth}
        \end{tabular}
    \label{tab:pce_power}
\end{table*}

\section{Conclusions}\label{conclusion}
Accurate short-term wind forecasts are indispensable for the optimal operation of (offshore) wind farms. We have proposed a physics-guided machine-learning-based forecasting model, called AIRU-WRF, which yields significant improvements, in terms of both point and probabilistic evaluations, over a wide array of forecasting benchmarks, thus testifying to the merit of AIRU-WRF to the OSW sector in the US and elsewhere. 

This work opens the door for several interesting future research avenues. For instance, meteorological models are often simultaneously run at multiple spatial resolutions (e.g., $3$ and $9$ km). AIRU-WRF can be extended to integrate the multi-scale physics represented by the nested NWP resolutions through a multi-resolution statistical modeling framework. In addition, extending the forecast horizon (e.g., to a day-ahead), varying the spatio-temporal resolution (e.g. turbine vs. farm-level), and extending AIRU-WRF to directly forecast wind power output, are all topics of ongoing research. 

\section*{Supplemental Material}
SM-1 is a video of the geopotential height over time. SM-2 is a video of AIRU-WRF's wind field forecast maps. {Both videos are appended to this article and are also accessible at \cite{airuwrfvideos}.} 

\section*{Acknowledgment}
{This work is supported in part by the U.S. National Science Foundation (ECCS-2114422) and in part by the National Offshore Wind Research \& Development Consortium (Project \# 192900-133). Funding for the RU-WRF model has been provided by the New Jersey
Board of Public Utilities.}



\bibliographystyle{elsarticle-num} 
\bibliography{references.bib}






\end{document}